\documentclass[manuscript]{emulateapj}

\newcommand{\myemail}{rorysmith274@gmail.com}

\slugcomment{Accepted to ApJ 16th November 2015}

\shorttitle{Disk of Satellites}
\shortauthors{Smith et al.}

\begin{document}


\title{A Formation Scenario for the Disk of Satellites: Accretion of Satellites during Mergers}


\author{Rory Smith\altaffilmark{1,2}, Pierre-Alain Duc\altaffilmark{2}, Frederic Bournaud\altaffilmark{2}, \and Sukyoung K. Yi\altaffilmark{1} }  
\altaffiltext{1}{Yonsei University, Graduate School of Earth System Sciences-Astronomy-Atmospheric Sciences, Yonsei-ro 50, Seoul 120-749, Republic of Korea; \myemail}
\altaffiltext{2}{Laboratoire AIM, Service d’astrophysique, CEA Saclay, Orme de Merisiers, Batiment 709, F-91191 Gif sur Yvette cedex, France; paduc@cea.fr}




\begin{abstract}
The Disk of Satellites (DoS) observed in the Andromeda galaxy is a thin and extended group of satellites, nearly perpendicular to the disk plane, that share a common direction of rotation about the centre of Andromeda. Although a DoS is also observed in the Milky Way galaxy, the prevalance of such structures in more distant galaxies remains controversial. Explanations for the formation of such DoSs vary widely from filamentary infall, or flattening due to the potential field from large scale structure, to galaxy interactions in a Mondian paradigm. Here we present an alternative scenario -- during a merger, a galaxy may bring its own satellite population when merging with another galaxy. We demonstrate how, under the correct circumstances, during the coalescence of the two galaxies, the satellite population can be spread into an extended, flattened structure, with a common direction of rotation about the merger remnant. We investigate the key parameters of the interaction, and the satellite population, that are required to form a DoS in this scenario.
\end{abstract}


\keywords{methods: N-body simulations -- galaxies: halos -- galaxies: interactions -- galaxies: kinematics and dynamics }



\section{Introduction}
For several decades it has been recognised that the luminous, classical Milky Way satellites exhibit an anisotropic spatial distribution \citep[e.g.][]{LyndenBell1976,LyndenBell1983,Majewski1994,Hartwick2000}. Subsequently it has been identified that they lie close to a virtual plane, known as the `Disk of Satellites' \citep[DoS,][]{Metz2008}. The virtual plane is highly inclined with respect to the stellar disk of the Milky Way. The members of the DoS have since been extended to include the ultrafaint dwarf spheroidals too \citep{Metz2009}. A common direction of rotation about the Milky Way centre was reported \citep{Metz2008}. Globular clusters and streams are also reported to be associated with the DoS \citep{Pawlowski2012}. Another DoS was also found about M31 \citep{Metz2009}. More recently, the DoS in M31 was found to be large ($\sim$200~kpc), thin ($\sim$20~kpc), rich ($\sim$15~members), and it also showed indications of rotation about the centre of M31 for 13 of the satellites \citep[`the vast thin plane,'][]{Ibata2013}. 

The discovery of a plane with such properties has spawned numerous follow up studies, observationally and theoretically. Planes with these properties are reported to be extremely rare in cosmological simulations \citep{Pawlowski2012,Bahl2014,Pawlowski2014} although more common if only the spatial distribution need be matched \citep{Wang2014}. While it is difficult to detect such rich planes of dwarfs in external galaxies, observational studies have attempted to compare numbers of diametrically opposed galaxy satellites in the low redshift Universe \citep{Ibata2014}. The results could suggest such planes are very common, and the observations may differ considerably from what would be expected in the standard model of galaxy formation. Although such conclusions remain controversial, and in fact \citet{Cautun2015} (see also \citealp{Phillips2015}) find good agreement between the standard model and observations. \citet{Cautun2015b} suggest that DoS are actually found in about 10$\%$ of $\Lambda$CDM halos, however their diverse range of properties means that DoS which exactly match the properties of the M31's DoS are rare, although the exact formation mechanism in this study is not revealed. A number of formation scenarios for DoSs have been proposed. 

In $\Lambda$CDM, a flattened distribution can arise from the infall of satellites along the spine of filaments \citep{Libeskind2005,Libeskind2011,Libeskind2014}, with a significant fraction of the satellites inheriting the spin of the host halo \citep{Libeskind2009,Lovell2011,Cautun2015}. Large scale structure and voids may also contribute to a local shear tensor that can produces satellite alignments \citep{Libeskind2015,Codis2015}.

Other scenarios outside of $\Lambda$CDM have been proposed, including dissaptive dark matter \citep{Randall2014}. It has also been suggested that the satellites of a DoS are actually tidal dwarf galaxies, formed in galaxy interactions in a Mondian Universe \citep{Kroupa2012}. This latter scenario notes that tidal dwarf galaxies naturally form along planes during galaxy interactions where the plane matches the original orbital plane of the interacting satellites \citep{Hammer2013}. Such a result does not require a Mondian Universe itself. However, due to their formation mechanism (gas streams dissipate and become clumpy), tidal dwarfs are found to collect negligible quantities of dark matter from their progenitor galaxies in a CDM scenario \citep{Bournaud2010}. Therefore if all of M31's DoSs were actually tidal dwarfs, then MOND would be additionally required to explain their observed high velocity dispersions \citep{Tollerud2012}. Furthermore, if all DoS dwarfs were formed simultaneously in a single tidal structure, a common star formation history might be expected (see \citet{Duc2014} for further discussion). Recently, \citet{Salomon2015} examined the intrinsic flattening and orientation of M31's dwarf spheroidals. While some are significantly elongated, no clear difference in the ellipticity distribution, nor the major axis alignment, are found between members and non-members of the DoS.

Here, we propose an alternative scenario. We consider a merger between a primary and a secondary galaxy. The secondary galaxy contains its own satellite population of dark matter dominated dwarfs. During the merger the secondary's dwarf satellites are thrown into a thin, extended, rotating plane, in much the same way as occurs during the formation of tidal dwarf galaxies. However, the advantage of this scenario is that these dwarfs are not tidal dwarfs, and contain dark matter from the outset. Therefore it is not necessary to appeal to MOND to describe the high velocity dispersion of the dwarfs internal dynamics. We conduct idealised N-body simulations of the merger between a primary and secondary halo, where the secondary halo contains a population of subhalos. We find that in this scenario, we can produce extended, thin, and rotating planes of satellites about the primary galaxy. We conduct a parameter study, varying orbital interaction properties, and dwarf satellite distribution and dynamics, to understand which parameters are key to producing an extended, thin, and rotating DoS in this scenario. For each parameter set, we quantify the resulting disk diameter, thickness, and the degree of satellite co-rotation. In $\oint$2, we describe our set-up and simulations, in $\oint$3 we show our standard model, and we conduct a parameter study, in $\oint$4 we discuss our results, and in $\oint$5 we draw conclusions.

\begin{figure}
\includegraphics[width=8.5cm]{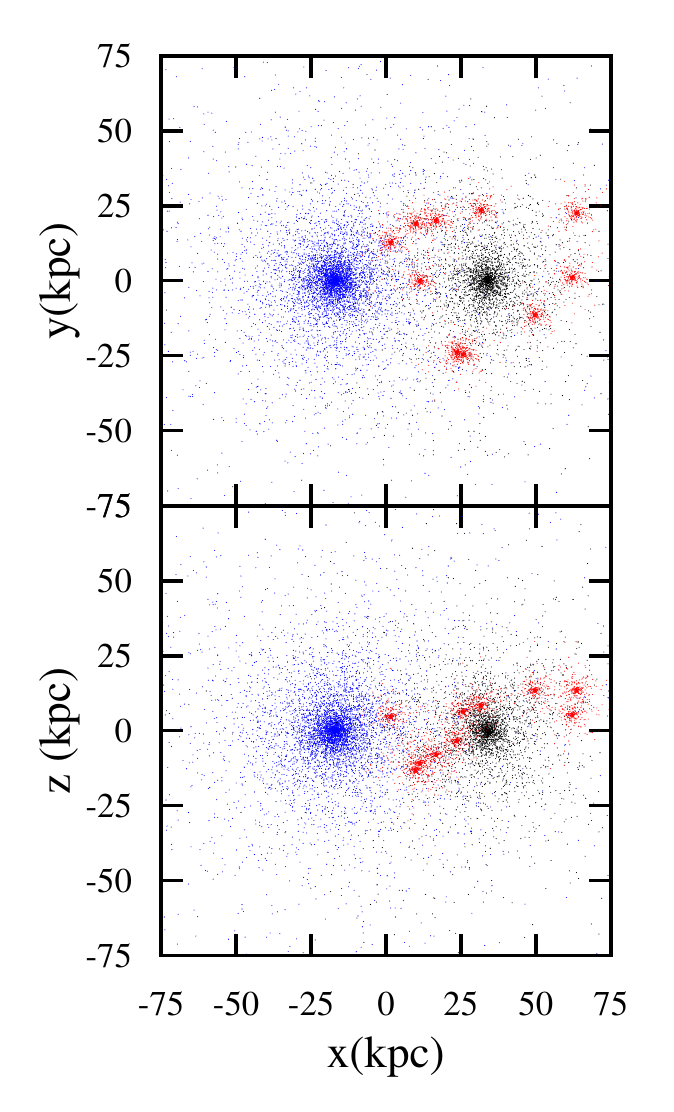}
\caption{$x$-$y$ projection (top panel) and $x$-$z$ projection (bottom panel) of the initial distribution of dark matter particles for the primary halo (blue points), secondary halo (black points), and dwarf halos (red points) of the standard model.}
\label{initialpos}
\end{figure}

\section{Method}
\subsection{Set up and numerical code}
\label{setupsection}
We set up idealised mergers between two dark matter halos. We use a Navarro, Frenk and White halo (NFW; \citealp{Navarro1996}) for all halos. The properties of the halos are derived based on cosmological halos formed at high redshift ($z=3$). We choose a 1:2 mass ratio merger which results in a major merger. We will return to the implications of this choice of mass ratio in the Discussion section. The more massive halo and the less massive halo will be described as the `primary' and `secondary' halo hereafter. The primary and secondary halo has a virial mass of $1 \times 10^{12}$~M$_\odot$ and $5 \times 10^{11}$~M$_\odot$, and a virial radius of 51.0~kpc and 40.0~kpc, respectively. NFW halos have a concentration parameter $c_{\rm{hal}}$, which is a measure of how centrally concentrated their dark matter distribution is. Both halos have concentration $c_{\rm{hal}}$=15, meaning each halo has 15 scalelengths within its virial radius. 

The secondary halo is initially placed at the virial radius of the primary halo. The secondary halo is given a velocity which is purely azimuthal with respect to the primary halo. The magnitude of the velocity vector is chosen to be a specified fraction $f_{\rm{circ}}$ of the circular velocity of the two body system, calculated using the classical two-body problem formula for two point masses. The value of $f_{\rm{circ}}$ is varied between simulation runs, from zero to one, to create mergers whose initial orbit while merging varies from highly radial to nearly circular. Regardless of our choice of $f_{\rm{circ}}$, the initial relative velocity between the primary and secondary halo quickly decreases due to dynamical friction. 

In the secondary halo, we place a population of ten subhalos which we will refer to as `dwarf halos'. Each dwarf halo has a virial mass of $4.0 \times 10^9$~M$_\odot$, virial radius 8.3 kpc, and concentration $c_{\rm{hal}}$=30. Satellite galaxies are typically not distributed isotropically within a host galaxy, and tend to show a preferential alignment \citep{Holmberg1969,Brainerd2005,Yang2006}. Therefore we consider different models, varying the spatial distribution, and dynamics of the dwarf halos within the secondary halo between models to try to identify the most important parameters. For example, in our `standard' model, it is assumed that the dwarf halos are initially distributed in a distribution that is preferentially flattened about the $x$-$y$ plane within the secondary halo. The $x$-$y$ plane is also the orbital plane of the interaction between the primary and secondary halo. The flattened distribution of dwarf halos extends to the virial radius of the secondary halo. We exclude dwarf halos within 20~kpc of the centre of the secondary halo, to avoid strong mass loss from dwarf halos due to internal tides. The dwarf halos are distributed in the z-direction, vertically out of the plane of the flattened distribution, with -20$<$$z$(kpc)$<$20. 

If the secondary halo were to evolve in isolation, the distribution of dwarf halos within it would be stable with time. This is accomplished by placing each halo on a stable, circular orbit within the potential of the secondary halo. The orbital plane of each individual dwarf is tilted with respect to the $x$-$y$ plane, with the degree of tilt being greater if the dwarf halo is initially further from the $x$-$y$ plane. For example, a dwarf that is initially at $z$=0~kpc moves on a circular orbit on the $x$-$y$ plane. Meanwhile, a dwarf that is initially at $z$=+20~kpc has an orbital plane that is tilted with respect to the $x$-$y$ plane such that it moves between z=+20~kpc and $z$=-20~kpc every half orbital period. In this way the dwarf distribution remains constant with time. In the standard model, all dwarfs are given a single, prograde direction of rotation about the centre of the secondary halo.

The distribution of dwarf halos in the initial conditions of the standard model shows preferential flattening, but is still quite thick (i.e. initially half as thick as it measures across in diameter, with a semi-major to semi-minor axis ratio $c$/$a$=0.5). The initial particle distribution in the $x$-$y$ and $x$-$z$ projection can be seen in Fig.~\ref{initialpos}. Blue points indicate primary halo particles, black points are secondary halo particles, and red points are particles belonging to the dwarf halo population. As we will demonstrate in the Sect.~\ref{stdmodel}, although we start with a distribution of dwarfs that is slightly flattened ($c$/$a$=0.5), the merger can form a final distribution of dwarfs that is much more strongly flattened ($c$/$a$=0.1-0.2). By considering distributions of dwarfs that are initially flattened, we can also study the survival of initially flattened satellite distributions to a galaxy merger. However, we will later relax the requirement for the dwarfs to be initially flattened, and/or to share a common direction of rotation, when we consider alternative models (see Sect.~\ref{altmodels}), and demonstrate that a DoS can be formed in these models as well.

All simulations are conducted with the adaptive mesh refinement code, {\sc{Ramses}} \citep{Teyssier2002}. Simulations are N-body models of just the dark matter component of the primary, secondary and dwarf galaxies. The cubic simulation volume is 400~kpc along one side. The maximum refinement level is 13, equivalent to a minimum cell size of $\sim$50~pc. We confirm that the code refines to the maximum level for dwarf halos, which ensures that we resolve their cusps down to below one third of their scalelength. Each dark matter particle has an equal mass of 4.0$\times$10$^5$~M$_\odot$, resulting in the total number of particles in the primary halo, secondary halo, and in a single dwarf halo being 2.5$\times$10$^6$, 1.25$\times$10$^6$, and 1.0$\times$10$^4$ respectively. We shift all simulations to the centre of mass frame to ensure the galaxy merger does not drift within the simulation volume. Each simulation takes approximately 24 hours, in parallelised runs on 256 cores.

\begin{figure*}
\includegraphics[width=16.3cm]{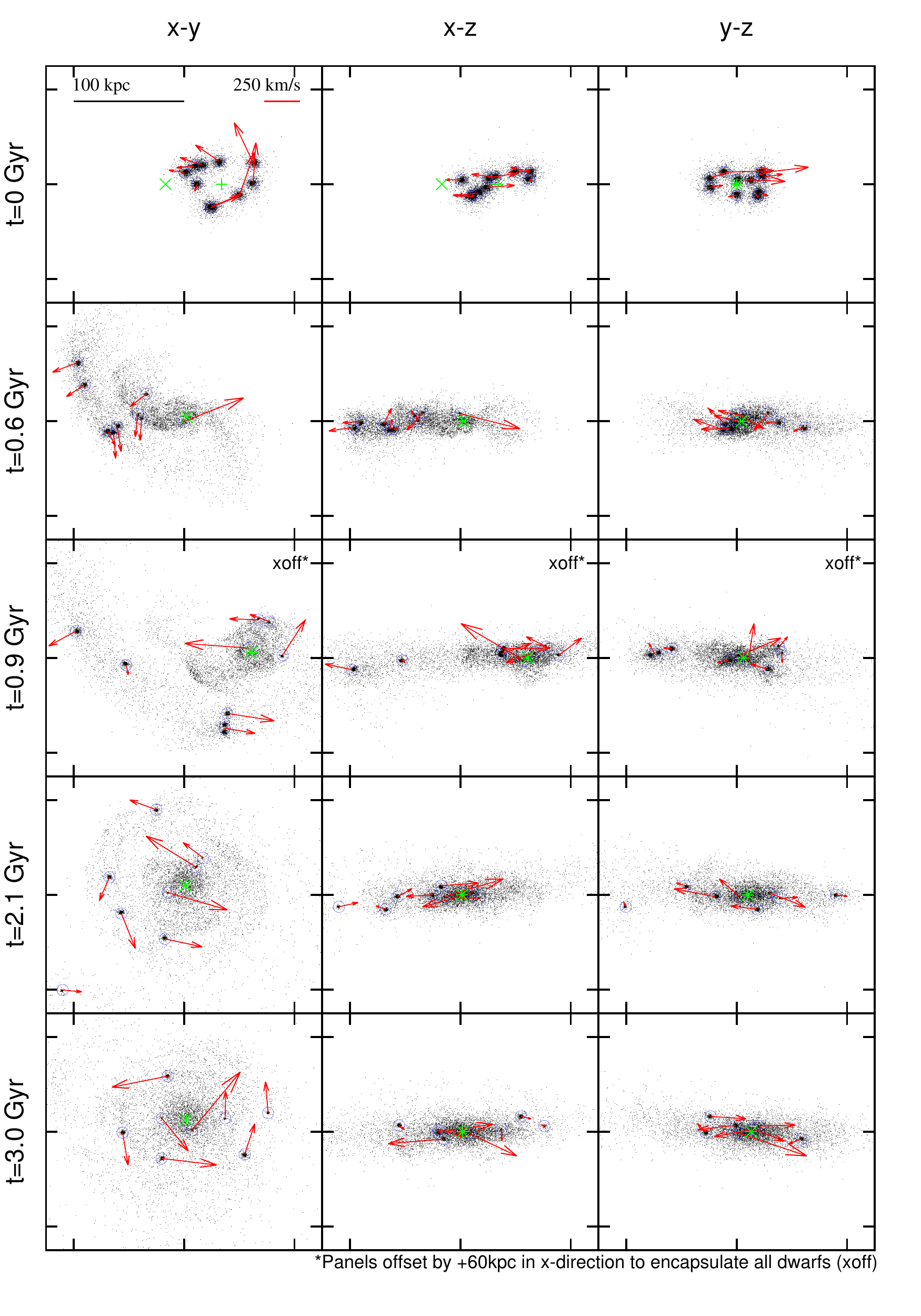}
\caption{$x$-$y$ projection (left column), $x$-$z$ projection (centre column), $y$-$z$ projection (right column) of only the dark matter particles associated with dwarf halos (black points). Each row indicates a different snapshot time as indicated on the far left, evolving from t=0.0~Gyr (top row), down to t=3.0~Gyr (bottom row). Blue circles highlight surviving halos. Red vector arrows show the velocity vector of each halo. A green cross and green plus symbol identifies the centre of the primary and secondary halo respectively (for t$>$0.6~Gyr, the primary and secondary halo are fully merged forming an asterisk symbol).In t=0.9~Gyr snapshots (third row; indicated with `xoff*' label), the model is shown offset by 60 kpc in the positive $x$-direction in order to allow a dwarf halo at large radius to remain within view.}
\label{standardevol}
\end{figure*}

\subsection{Time evolving properties of DoS}
Typically the merger between the primary and secondary halo is completed within approximately half a gigayear, as seen in other similar mass ratio merger simulations \cite[e.g.][]{Ji2014}. Therefore we conduct all simulations for a total time of 3~Gyr, in order to give us time to follow the evolution of the DoS for a few gigayears beyond the merger. Each dwarf halo is identified by first choosing all particles above a number density limit (in practice we choose 10 particles within a 0.5~kpc radius sphere). Then, particles are associated with a particular dwarf halo if they lie within 2.5~kpc (roughly one third of the initial dwarf halo virial radius)  of each other. We then calculate the number of dwarfs, and their average position and velocity, which we use to measure the DoS properties; size and thickness of the DoS, and fraction of dwarf halos with the same direction of rotation, as a function of time. 

In the following, the $x$ and $y$ axis are aligned with the plane of the DoS, while the $z$ axis is perpendicular to the plane. We measure the thickness of the DoS ($\Delta$$z$) as the maximum $z$-range of the dwarfs. We measure the diameter of the DoS ($D_{\rm{avg}}$) by taking the average of the maximum $x$-range and $y$-range of the plane. We define the semi-major to semi-minor axis ratio as $c$/$a$=$\Delta$$z$/$D_{\rm{avg}}$. Because we choose this definition, it is possible for a model to have $c$/$a$$>$1, should $\Delta$$z$ become larger than $D_{\rm{avg}}$. However, in practice all models maintain $c$/$a$$<$1,  with the exception of a single model (the shell model) that surpasses this value for a small fraction of the simulation time.

The number of surviving dwarf halos is measured, assuming dwarf halos are destroyed once they have less than 5$\%$ of their initial mass remaining. This choice ensures the removal of galaxies that are on the verge of complete disruption. Such galaxies would be heavily stripped of their baryons, and so difficult to detect. 

Finally, we measure the fraction of dwarfs orbiting with the same, prograde direction about the the galaxy centre, by measuring the normal vector to plane in which each dwarf halo moves, and testing if its $z$ component is positive or negative. 

The properties of the DoS can vary significantly over the duration of the simulation. Therefore we measure the time-averaged properties of the DoS, from t=0.5~Gyr (shortly after the merger) until the end of the simulation. The time-averaged properties of all the models are presented in Tab. \ref{tavgpropstab}.

\section{Results}
\subsection{The standard model}
\label{stdmodel}
\subsubsection{Snapshots of the standard model evolution}
In Fig.~\ref{standardevol}, we show snapshots of the evolution of the position and dynamics of the dwarf halos during the merger. In this case, we choose the initial azimuthal velocity of the secondary halo, with respect to the primary halo, to be 0.75 times the circular velocity of the two-body system ($f_{\rm{circ}}$=0.75). The circular velocity is calculated using the classical solution to the gravitational two-body problem, where two point masses orbit about the so-called `reduced-mass'. For all our models, the circular velocity is 260.7~km/s. We will consider the effects of varying this choice of initial relative velocity in the following section.

In Fig.~\ref{standardevol}, we show the position of dark matter particles (black points) associated with dwarf halos only. The primary and secondary halo particles are neglected for clarity, however we indicate the centres of their halos by a green cross and plus symbol respectively. Each row indicates a different snapshot times (snapshot time is shown on the left of the row), with time evolving from top to bottom row. We show the particles viewed along three different projections; $x$-$y$ is the left column, $x$-$z$ is the centre column, $y$-$z$ is the right column. Blue circles highlight clumps of particles that are identified as surviving halos, and red vector arrows show the velocity vectors of these clumps. Velocity vectors are calculated with respect to the frame of reference of the simulation volume. 

At t=0~Gyr (row~1), all the dwarfs share a common, prograde direction of rotation about the centre of the secondary halo. Viewed in the frame of reference of the secondary halo, their orbits would appear near circular. However, it does not appear this way in the upper-left panel because the velocity vectors are calculated with respect to the frame of reference of the simulation volume (the centre-of-mass frame). Nevertheless, when the thick distribution of satellites is viewed edge on (e.g. upper-centre and upper-right panel), the velocity vectors can be seen to be confined to the plane of the disk.

By t=0.6~Gyr (row~2), the primary and secondary galaxy are fully merged (the green cross and plus symbol overlap to form an asterisk symbol). Some dwarfs are unbound from the secondary halo by the tidal interaction with the primary halo, and several are slung outwards towards the left of the panel (in the negative $x$-direction. However, viewed edge-on (centre and right panel), it can be seen that the dwarfs are slung out while remaining in the orbital plane that the primary and secondary halo merged on (the $x$-$y$ plane).

By t=0.9~Gyr (row~3), some of those dwarfs that were slung out have reached large radii ($\sim$180~kpc). As a result, the t=0.9~Gyr panels are shown offset by 60~kpc in the positive $x$-direction, in order to keep all dwarfs in view. The $x$-$y$ projection (left panel) reveals that all dwarfs are rotating in the same direction about the centre of the merger remnant. However the DoS is still lopsided at this early stage, as the dwarfs were slung out preferentially in one direction. Viewed edge-on (centre and right panel), most of the velocity vectors are shown to remain within the original orbital plane of the primary and secondary halo's interactions. This indicates that the thickness of the DoS will be long-lived in our model, and is not a by-chance incident when the dwarfs have fortunately lined up at one instant. Only one dwarf, which has strayed close to the centre of the merger remnant and been scattered, shows a velocity vector at a steep angle with respect to the plane of the DoS. 

At t=2.1~Gyr (row~4), the DoS is much more symmetrical in the $x$-$y$ plane (left panel), and it can be seen that the DoS remains very extended, measuring $\sim$150~kpc in diameter. Meanwhile, viewed edge-on (centre and right panel), all dwarfs remain with approximately the same spread vertically out of the plane of the DoS (in the z-direction) as they had in the thick distribution they were initialised in (compare to the centre and right panel of the upper row). This means the semi-major to semi-minor axis ratio of the dwarf distribution has been decreased from an initial value of $c$/$a$=0.5 to a value of $c$/$a$$\sim$0.15 -- a thin DoS, that is a factor of more than three times thinner than the initial dwarf distribution about the centre of the secondary halo. 

Finally, at t=3.0~Gyr, the DoS can be seen to remain in place at $\sim$2.5~Gyr since the merger between the primary and secondary halo occurred. The DoS is now much more axi-symmetric about the merger remnant, and vector arrows indicate a single direction of rotation for all surviving dwarfs. Of the initial ten dwarf halos, two have been tidally destroyed. Therefore the DoS consists of the eight remaining dwarfs. The spread in the $z$-direction is nearly unchanged over the entire duration of the simulation, but the diameter of the DoS is greatly expanded. Thus the semi-major to semi-minor axis ratio decreases from an initial value of $c$/$a$=0.5 to a final value of $c$/$a$=0.2, almost purely as a result of an increase in $a$, with little change in $c$.

\subsubsection{Dependency on circularity of initial orbit}
In the previous section, we considered the standard model when the initial azimuthal velocity between the primary and secondary halo was fixed to be 0.75 times the two-body circular velocity ($f_{\rm{circ}}$=0.75). In the following section we consider a range of bound elliptic orbits, by varying the initial azimuthal velocity from $f_{\rm{circ}}$=0.3-1.0. In this way we investigate how the initial circularity of the orbit impacts on the properties of the DoS that form.

As DoS are formed from small numbers of dwarf halos in our models, we conduct 5 realisations of every $f_{\rm{circ}}$ value. In each realisation, we use a random number generator to form a new, and unique distribution of satellites about the centre of the secondary halo. In fact, each random realisation has the same flattened distribution as in the standard model, but the sampling of that flattened distribution varies between realisations, due to the low numbers of dwarf halos used to sample it. Therefore, to better overcome low-N statistics, for every $f_{\rm{circ}}$ value we calculate a mean and standard deviation of each DoS parameter.

\begin{figure}
\includegraphics[width=9.cm]{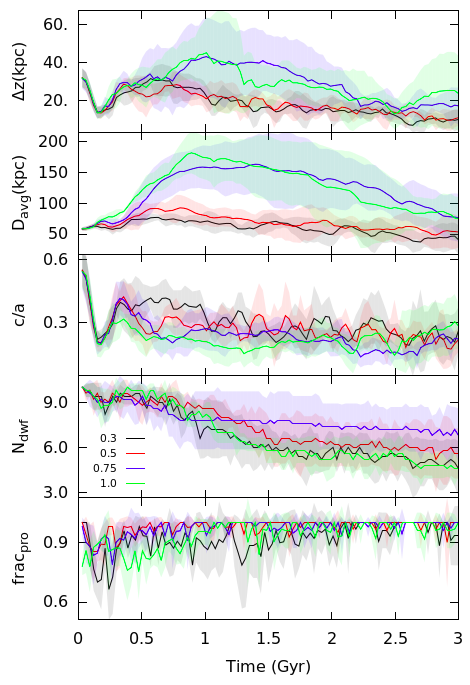}
\caption{Time evolution of the DoS properties for different choices of initial relative tangential velocity between the primary and secondary halo. Line colour indicates the value of the tangential velocity in units of the circular velocity, calculated using the classical solution for a two-body system. Five models of each tangential velocity are simulated, each with a different random realisations of the dwarf distribution, in order to better overcome low-N statistics. The solid line indicates the mean of the 5 models, and the shaded area indicates their standard deviation. From top to bottom panel, we consider the evolution of the following DoS properties; thickness ($\Delta$z), diameter ($D_{\rm{avg}}$), semi-major to semi-minor axis ratio ($c$/$a$), number of surviving dwarfs ($N_{\rm{dwf}}$), and fraction of dwarfs that rotate in the prograde direction (${\rm{frac}}_{\rm{pro}}$).}
\label{planeevol_orb}
\end{figure} 

In Fig.~\ref{planeevol_orb}, we show how the properties of the DoS evolve with time for each value of $f_{\rm{circ}}$ considered. The mean value of the five random realisations is shown as a solid line, where as the shaded region surrounding each line indicates the standard deviation.

The figure shows that the thickness and diameter of the DoS depend on the $f_{\rm{circ}}$ value used (top and second panel). DoS with a large diameter ($>$100~kpc) form shortly after merging, and last approximately 2~Gyr, when $f_{\rm{circ}}$ is 0.75 or greater. We note that the strong effect of dynamical friction quickly decays the orbit of secondary halo into the primary halo for all the models. However for values of $f_{\rm{circ}}$$>$0.75, the secondary halo spirals into the centre of the primary halo, while maintaining a roughly circular orbit, before merging at around t=0.6~Gyr. This roughly circular orbit is beneficial to forming the DoS in two ways. Firstly, if the dwarf system were to have a very radial orbit, it could plunge through the centre of potential well of the primary galaxy. This could result in scattering of the dwarfs off the minimum of the primary galaxy's potential well, destroying any opportunity for a thin DoS. Secondly, the more circular orbit is beneficial for angular frequency matching between the dwarfs orbiting about the secondary, and the secondary halo orbiting about the primary halo. As a result, the dwarfs are more effectively slung outwards to large radii during the interaction. This occurs for both models with high $f_{\rm{circ}}$ values ($f_{\rm{circ}}$=0.75 and $f_{\rm{circ}}$=1.0), but fails to occur for the models with $f_{\rm{circ}}$=0.3 or $f_{\rm{circ}}$=0.5. The diameter of the DoS evolves significantly over the duration of the simulations. However, the time-averaged diameters of the $f_{\rm{circ}}$=0.75 and 1.0 models are significantly larger than that of the $f_{\rm{circ}}$=0.3 and 0.5 models (see Tab. \ref{tavgpropstab}).

The third panel shows that for all of the $f_{\rm{circ}}$ values considered, the semi-major to semi-minor axis ratio ($c$/$a$) reduces from its initial value. This occurs due to decreasing DoS thickness combined with increasing disk diameter. The time-averaged $c$/$a$ ratio is between 0.21-0.29 for all $f_{\rm{circ}}$ models, compared to an initial value of 0.50$\pm$0.13.

The fourth panel of Fig.~\ref{planeevol_orb} shows that most models lose 3-6 dwarf halos due to tidal disruption by the end of the simulation. The time-averaged number of dwarfs varies from 6.3-7.6 between $f_{\rm{circ}}$ models. There is no clear dependence on $f_{\rm{circ}}$, given the size of the scatter in the number of surviving dwarf halos between random realisations.

Given that the DoS are formed by small numbers of dwarf satellites, it is important to understand the significance of our measured $c$/$a$ values. To do so, we set up a uniform sphere of dwarf halos, with only 5 or 10 dwarfs in total. We use a random number generator to produce 1000 random, and unique realisations. For each uniform sphere we measure the semi-major to semi-minor axis ratio ($c$/$a$) in exactly the same way as it is done for our DoS models. We find that, in most cases, models have $c$/$a$ close to 1, as might be expected for a uniform sphere. In fact, for uniform spheres consisting of 10 halos, 85$\%$ have 0.75$<$$c$/$a$$<$1.25. The third panel of Fig.~\ref{planeevol_orb} shows that most of the DoS that form in our merger models have $c$/$a$$<$0.5. For uniform spheres consisting of 10 halos, only 0.3$\%$ of 1000 random realisations have $c$/$a$$<$0.5. Thus a measured $c$/$a$$<$0.5 is of high significance. For uniform spheres consisting of 5 halos, the probability is higher. 5.3$\%$ of 1000 random realisations have $c$/$a$$<$0.5. Thus a measured $c$/$a$$<$0.5 is still of high significance.

The bottom panel of Fig.~\ref{planeevol_orb} shows that a common prograde direction of rotation is shared by all the surviving members of the DoS, even when $f_{\rm{circ}}$ is as low as 0.3. The time-averaged value of the fraction of dwarfs with prograde rotation ($frac_{\rm{pro}}$) is $>$0.93 for all $f_{\rm{circ}}$ models.

\begin{figure}
\includegraphics[width=9.5cm]{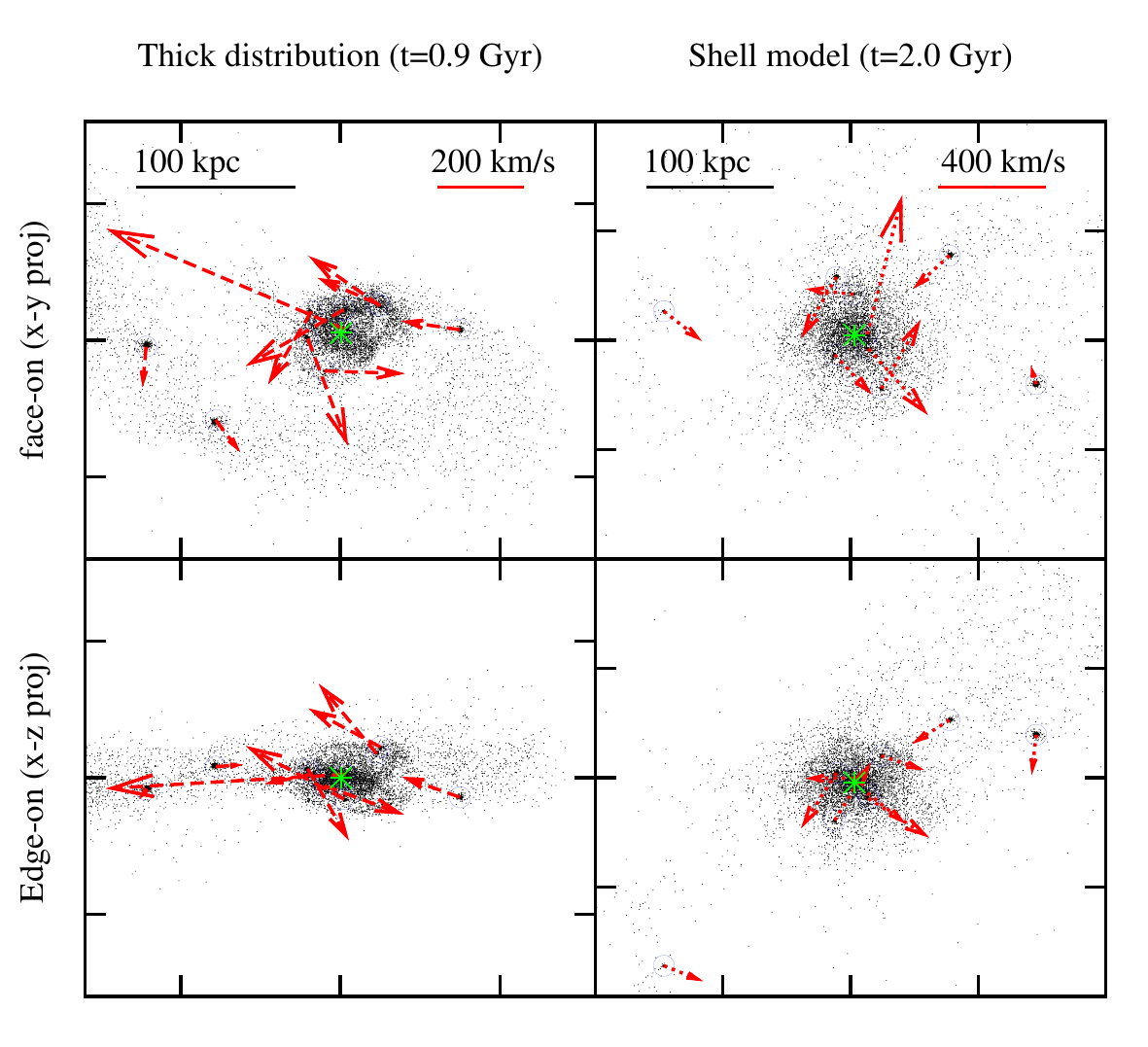}
\caption{A single snapshot of the DoS formed in the thick distribution model (left column), and the shell model (right column). The time of the snapshot is shown in the title of the column. See text in Sect.~\ref{altmodels} for further details of individual models. Face-on ($x$-$y$) projection is shown in the upper row, edge-on ($x$-$z$) projection is shown in the lower row. Symbols, and vectors are as in the caption of Fig.~\ref{standardevol}. Vector line style distinguishes different models, and the line style matches that in Fig.~\ref{planeevol_mod}.}
\label{othermodels}
\end{figure} 

\subsection{Other models: varying the distribution of the dwarf halos}
\label{altmodels}
We consider several alternative models to the standard model, varying the distribution and dynamics of dwarf halos within the secondary halo. Each model's properties is described in turn below, and the DoS that is produced is shown in Fig.~\ref{othermodels} along two projections; face-on ($x$-$y$ projection) and edge-on ($x$-$z$ projection). We vary the time at which the DoS is shown in Fig.~\ref{othermodels} between the models, in order to find moments when the DoS is large and thin. However, to remedy this, in Fig.~\ref{planeevol_mod} we present the time evolution of the DoS parameters. To aid comparison, we use the same line styles for the vector arrows in Fig.~\ref{othermodels} as for the curves in Fig.~\ref{planeevol_mod}. In addition, the instants of the snapshots in Fig.~\ref{othermodels} are indicated with filled symbols on Fig.~\ref{planeevol_mod}.

\subsubsection{The thick distribution model}
In the `thick distribution' model, we follow the same set-up procedure as was used for the standard model in order to produce a stable, flattened distribution of dwarfs (see Sect. \ref{setupsection}). However, unlike in the standard model, we reduce the inner and outer radii of dwarfs about the centre of the secondary halo to 5~kpc and 20~kpc, while maintaining the same distribution in the $z$-direction (20$<$$z$(kpc)$<$20). As a result the dwarf distribution is thicker than the standard model, as it has a smaller diameter, while maintaining the same vertical thickness. After randomly selecting dwarf positions within these ranges, the thick distribution model has an initial semi-major to semi-minor axis ratio $c$/$a$=0.75, and so is only weakly flattened. 

This model allows us to test the importance of having a flattened distribution for forming a DoS. We find that a relative velocity of 0.75 times the circular velocity of the two-body system ($f_{\rm{circ}}$=0.75) can well produce a DoS (the same initial relative velocity in the standard model). In the first panel of Fig.~\ref{othermodels} the DoS is shown at time=0.9~Gyr. The face-on view reveals a DoS with diameter $\sim$150~kpc, consisting of nine of the original ten dwarf halos. All dwarfs rotate in the same direction, although two move on fairly radial orbits. This indicates that {\it{the dwarfs do not need to be in a distribution that is highly flattened to form a DoS in this way}} as, even when the spatial distribution is only slightly flattened, we can still form a DoS. The DoS that results is thin, with semi-major to semi-minor axis ratio of $c$/$a$$\sim$0.2. 

\begin{figure}
\includegraphics[width=9.5cm]{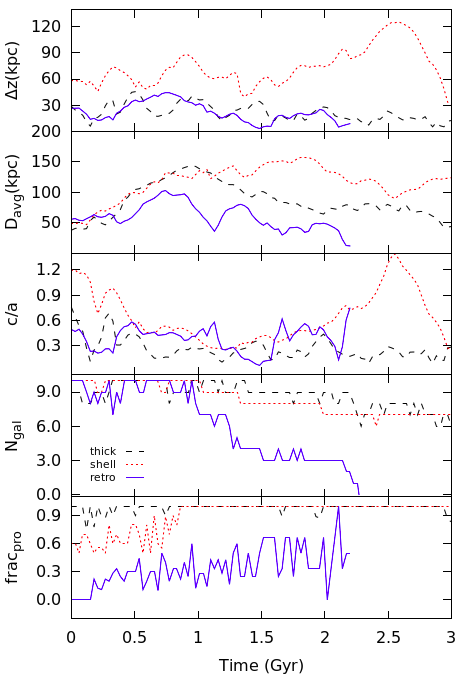}
\caption{Time evolution of the DoS properties for each of the alternative models considered in Sect. \ref{altmodels}. Line style indicates the model (see legend). Line style is chosen to match those given for each model in Fig.~\ref{othermodels} for ease of comparison. From top panel to bottom, we consider the evolution of the following DoS properties; thickness ($\Delta$z), diameter ($D_{\rm{avg}}$), semi-major to semi-minor axis ratio ($c$/$a$), number of dwarfs ($N_{\rm{dwf}}$), and fraction of dwarfs that rotate in the prograde direction (${\rm{frac}}_{\rm{pro}}$).}
\label{planeevol_mod}
\end{figure}

Fig.~\ref{planeevol_mod} shows that the thickness of the DoS in this model does not evolve substantially throughout the duration of the simulation, but the disk diameter peaks around the time of the snapshot in Fig.~\ref{othermodels}. The $c$/$a$ ratio quickly decreases following the merger at t=0.5~Myr, and remains at a value near 0.2 thereafter. The number of surviving dwarfs steadily decreases with time, with 6 dwarfs halos remaining at the end of the simulation. However all surviving dwarfs share the same direction of rotation. The time-averaged $c$/$a$ ratio, number of dwarfs, and fraction of prograde dwarfs is 0.22, 8.6 and 0.99 respectively (see Tab. \ref{tavgpropstab}).

\subsubsection{The shell model}
In the `shell' model, the dwarfs are initially randomly distributed on a spherical shell at a radius of 30~kpc from the centre of the secondary halo. Thus the initial flattening of the distribution of dwarfs has been entirely abandoned. Each dwarf has the circular velocity of the secondary halo, but their velocity vector has random direction on the shell. With these models, we can test the importance of the degree of anisotropy of the dwarf population for forming a DoS. We find that a relative velocity of 1.4 times the circular velocity of the two-body system ($f_{\rm{circ}}$=1.4) produces a reasonable DoS (this is a 40$\%$ larger initial relative velocity than considered in the standard model). In the second panel of Fig.~\ref{othermodels} the DoS is shown at time=2.0~Gyr. Three dwarfs have been tidally destroyed, thus the DoS consists of the 7 remaining dwarf halos. The face-on view shows a DoS with diameter $\sim$150~kpc. As with previous models, all dwarf halos share the same direction of rotation. An edge-on view reveals that the DoS is not in the plane of the orbital interaction of the primary and secondary halo. Furthermore, the DoS is not as thin as in the models where we begin with a thick distribution of dwarfs in the secondary halo, and their velocity vectors are not well aligned within the DoS. This strongly suggests that, {\it{to produce a thin DoS, it is necessary for the dwarf halos to have a small range of $z$ prior to the merger}}, where $z$ is the direction perpendicular to the plane of interaction between the primary and secondary halo. For example, when we varied the radial distribution, without varying the $z$-range (i.e. between the standard model and the thick distribution model), we could still produce a thin DoS. However, when we consider a shell model, which has a much larger range of $z$, we get a thicker DoS. In addition, if we test the standard model, but with the plane of the flattened distribution initially perpendicular to the orbital plane, we find a thin DoS is not formed. Finally, when we repeat the shell model experiment, but this time randomising the position of the dwarfs again, we get a much thicker DoS. This indicates that this DoS is a more by-chance occurrence, where the dwarf halos within it had, by-chance, a small range of $z$ at a key moment during the merger. Because it was by-chance, the velocity vectors were not well aligned with the plane of interaction between the primary and secondary halo, and the result is that the dwarfs in the DoS have velocity vectors which are not well aligned with the plane of the DoS.

\begin{table}
\centering
\begin{tabular}{p{1.4cm}|c|c|c|c|c|p{1.cm}}
Model & ($c/a$)$_{\rm{i}}$ & $\Delta$$z$(kpc)& $D$$_{\rm{avg}}$(kpc) & $c/a$ & $N$$_{\rm{dwf}}$ & $frac$$_{\rm{pro}}$  \\
\hline
$f$$_{\rm{circ}}$=0.3 & 0.50 & 17.1 & 59.1 & 0.29 & 6.3 & 0.93 \\
$f$$_{\rm{circ}}$=0.5 & 0.50 &17.6 & 68.9 & 0.26 & 7.0 & 0.99 \\
$f$$_{\rm{circ}}$=0.75$^\star$& 0.50 &29.3 & 132.5& 0.22 & 7.6 & 0.98 \\
$f$$_{\rm{circ}}$=1.0 & 0.50 &28.3 & 131.0& 0.21 & 6.3 & 0.95 \\
thick                 & 0.75 &20.7 & 91.8 & 0.22 & 8.6 & 0.99 \\
shell                 & 1.20 &75.7 & 126.5& 0.58 & 8.1 & 0.96 \\
retro                 & 0.50 &22.3 & 60.2 & 0.38 & 5.7 & 0.40 \\
\end{tabular}
\caption{Time-averaged values of DoS properties measured after the galaxies have merged (for t$>$0.5 Gyr). From left to right column; model name, initial semi-major to semi-minor axis ratio ($c/a$)$_{\rm{i}}$, thickness ($\Delta$$z$), diameter ($D$$_{\rm{avg}}$), semi-major to semi-minor axis ratio ($c$/$a$), number of dwarfs ($N$$_{\rm{dwf}}$), and fraction of dwarfs with prograde motion ($frac$$_{\rm{pro}}$). ($^\star$The standard model is one of the five $f$$_{\rm{circ}}$=0.75 models used to calculate the time-averaged properties of the DoS.)}
\label{tavgpropstab}
\end{table}

The first and second panel of Fig.~\ref{planeevol_mod} show that the shell model DoS actually has a comparable diameter to the standard model. However it is thicker, and so has a larger semi-major to semi-minor axis ratio $c$/$a$. The time-averaged value of $c$/$a$ is 0.58, which is considerably larger than all other models (see Tab. \ref{tavgpropstab}). Furthermore, as the velocity vectors are not well aligned with the DoS, the thickness of the DoS varies more with time than in the standard and thick distribution model. Dwarf halos are destroyed at a similar rate as in the standard model, with seven out of ten remaining at the end of the simulation. Once more, all surviving dwarf halos share the same, prograde direction of rotation. The time-averaged number of dwarfs, and fraction of prograde dwarfs is 8.1, and 0.96 respectively.

\subsubsection{The retrograde rotation model}
In our standard and thick distribution model, the dwarf halo population were set up such that all had the same prograde direction of rotation. In the `retrograde rotation' model, we consider a model which is identical to the standard model, except we choose the rotation direction of the dwarf halos to be retrograde about the centre of the secondary halo. Thus we can directly compare the standard model and retrograde model. We find that the behaviour and dynamics of the prograde and retrograde dwarfs is very different. The standard (prograde) model produces a large diameter disk with a small semi-major to semi-minor axis ratio (the time averaged diameter is 131~kpc). An average of 7 dwarfs populate the DoS by the end of the simulation. In contrast, the retrograde dwarf model produces a much smaller diameter disk (see second panel of Fig.~\ref{planeevol_mod}). The time-averaged diameter is only 60.2~kpc (see Tab.~\ref{tavgpropstab}). The third panel of Fig.~\ref{planeevol_mod} reveals that all the retrograde dwarfs are tidally destroyed by the end of the simulation. This is because the retrograde dwarfs end up with highly radial motions, and so experience significantly more damaging tides than for the more circular orbits of the initially prograde dwarfs. 

This clear difference in behaviour occurs due to angular frequency matching between the dwarfs orbiting about the secondary halo, and the secondary halo orbiting about the primary halo. As a result, the velocities of the dwarfs about the secondary halo are similar to the interaction velocity between the primary and secondary halo. Therefore those dwarfs on retrograde orbits have effectively a near zero azimuthal velocity with respect to the merger remnant, and fall into its centre on very radial orbits. The lower panel of Fig.~\ref{planeevol_mod} shows that the retrograde dwarfs are converted onto highly radial orbits, resulting in a large amount of scatter in the fraction of objects sharing a prograde direction of rotation. We note that the spatial measurements cannot be calculated once the dwarfs are destroyed, hence the lines abruptly halt after t=2.2~Gyr.

This result demonstrates that it is not a key requirement that the initial dwarf halo population have a common direction of rotation about the centre of the secondary halo in order to form a DoS. Given a secondary halo with a mix of prograde and retrograde orbit dwarfs, angular frequency matching naturally allows the dwarfs on prograde orbits to form the rotating DoS. Meanwhile dwarfs on retrograde orbits are automatically filtered out, and finish with very radial orbits that result in stronger tidal mass loss. On long-time scales, this causes preferential destruction of the retrograde dwarfs. This could potentially result in a larger fraction of the dwarf population having the same direction of rotation within the DoS, as those that do not are preferentially tidally destroyed.

\section{Discussion}

By comparing between the models, we can identify which parameters are important to form a DoS in our scenario. We find that the radial extension of the dwarfs about the secondary halo is not important, as long as the dwarfs initially have a small $z$-range, where $z$ is the direction perpendicular to the plane of the orbital interaction of the primary and secondary halo. In other words, the initial distribution of dwarfs need not be significantly flattened (e.g. $c$/$a$=0.75), as long as $c$ is small. In the process of merging the dwarfs are slung out to larger radius, while decreasing their $z$-range. Thus the final DoS has a much smaller semi-major to semi-minor ratio ($c$/$a$), but this arises by increasing $a$, while simulataneously reducing $c$. Is the DoS we produce in our scenario thin enough to match the observed DoS about the MW and M31? Considering the classical dwarfs and ultrafaint dwarfs in the MW (see \citealp{Pawlowski2012}), the DoS is roughly $\sim$300~kpc across, by $\sim$100~kpc thick, therefore has a $c$/$a$ ratio of approximately one third. This is thicker than in our standard and thick distribution model where we produce $c$/$a$$\sim$0.15-0.25, and a time-averaged value of $c$/$a$=0.22 for both the standard and thick distribution model. The DoS of the M31, however, is much thinner (see \citealp{Ibata2013}) with a $c$/$a$$<$0.05. To produce such a thin DoS in our scenario would require a very compact $z$-range ($<$15~kpc) of the dwarf halos prior to the collision. However \citet{Cautun2015b} find that the DoS produced in $\Lambda$CDM simulations come with a wide range of properties. Therefore we expect that our scenario could potentially be an important source of DoSs in cosmological simulations such as those considered in \citet{Cautun2015b}. Indeed a range of DoS properties would be expected due to variations in the properties of the merger (e.g. redshift, mass ratio, orbital circularity, etc). How often could dwarfs form in nature with a narrow range of $z$ is less clear. In the `Holmberg effect' \citep{Holmberg1969} satellites were observed to be preferentially aligned with the minor axis of their host galaxies. However recently this has been refuted with \citet{Brainerd2005} and \citet{Yang2006}, who find there is preferential alignment with the {\it{major}}-axis of the host galaxies, especially at small projected radii to the host. Curiously, the alignment is found to be strongest in red hosts with red satellites \citep{Yang2006}.

Another potentially important parameter is the direction of the velocity vectors of the dwarfs with respect to the orbital interaction plane of the primary and secondary halo. In the shell model, the velocity vectors were not well aligned with the orbital plane -- the result is a thicker DoS, and the satellites within the DoS have velocity vectors that do no lie closely within the plane of the DoS. This causes greater time evolution of the DoS compared to the the standard model and thick distribution model. In these two latter models, we see little indication for the DoS thickening with time over the duration of the simulations ($\sim$3~Gyr). One potential source of heating could be scattering of dwarfs off the centre of the merger remnant, but this appears to be rare, thanks in part to fairly circularised orbits. To investigate the longevity of our DoS, we double the duration of the standard model simulation from three to six gigayears. We see no indication of a significant change in the properties of the DoS over the additional three gigayears. Hence the DoS remains stable for at least 5.5~Gyrs since its formation, with little indication that this would change if the models would be continued for longer. However, we note that our models are evolved in total isolation. DoS that form in a cosmological context could potentially be heated by subsequent mergers, and by tidal interactions with halos of other galaxies and groups. Therefore the thickening of the DoS in our models with time is probably a best-case scenario.

A final important parameter for producing DoS was that our models preferred the merger between the primary and secondary halo to occur on a fairly circularised orbit (see Fig.~\ref{planeevol_orb}). How often are such orbits expected to occur? \citet{Wetzel2010} studied the orbits on satellites on first infall into host halos in cosmological simulations. Satellites that infall into lower mass systems have more circular orbits. There is also a redshift dependence, with more recent infalls being more circular. Therefore our scenario may occur more frequently in galaxy mergers when the primary halo is lower mass, and for more recent mergers.

In our models, we only consider the dark matter component of the galaxies involved in the mergers. How would the inclusion of baryons influence our results? The principle mechanism by which our DoSs are formed is governed by dynamical friction, and evolving potential fields. Therefore it is unlikely that the addition of a hydrodynamic forces would be of much consequence for the DoS formation. Assuming the stellar mass is a small fraction of the total mass, as favoured by abundance matching \citep{Guo2010}, it is unlikely that the inclusion of baryons would change the net potential fields experienced by the DoS significantly either. However, the presence of a thin stellar disk could result in more tidal destruction of dwarf halos prior to the merger. Furthermore, the flattened, axi-symmetric potential well of a thin stellar disk could enhance orbital precession, if there is some degree of misalignment between the stellar disk and DoS, which could lead to more DoS thickening. Our scenario has an important consequence for stars in the dwarfs of the DoS. As our dwarf galaxies are not formed during the merger, their stellar populations would be expected to be typical of other dwarf galaxies that were not involved in the merger. This is distinct from the scenario where DoS dwarfs are actually tidal dwarf galaxies \citep{Kroupa2012}, and so would be expected to have distinct signatures of the epoch of their formation in their stellar populations.

We note that we have so far only considered the case where a dwarf population exists within the secondary halo. It is reasonable to expect that a dwarf population would also exist in the primary halo. For this reason we construct a model which is identical to that of the shell model, except we instead place the dwarf population within the primary halo prior to the collision. We find that tidal torques of the merger are less effective at slinging dwarfs out of the primary halo into a DoS. This is because the primary halo dominates over the mass of the secondary halo, and so is relatively weakly influenced by the torques from the secondary halo. However we note that the primary halo's dwarfs do pick up a weak overall rotation in the same direction as occurs when the satellites are in the secondary halo instead. We also construct an additional model where the primary and secondary halo are equal mass. In both the primary and the secondary halo we place a dwarf distribution like that of the standard model. We consider the case where both dwarf populations have prograde motion. With equal mass, both of the halos can torque each other's population of dwarfs, and the result is that both sets of dwarfs join the DoS. Therefore we can see that, when both halos contain dwarf satellites, the end result is rather sensitive to the mass ratio in the merger. When the mass ratio is equal, the dwarf population in both systems can form the DoS. However if the mass ratio is 1:2 (or even more minor), only the dwarfs from the secondary halo can form the DoS. As such, the final system would consist of a DoS surrounded by the more spherical distribution of dwarfs from the primary halo, which have a slow rotation in the same direction as the DoS.

In this study, we have mainly focussed on a single merger mass ratio (1:2). A change in mass ratio between the primary and secondary halo is important for the dynamical friction timescale, with higher mass ratios (e.g. 1:10) resulting in longer merging timescales ($>$0.5~Gyr). Nevertheless, this doesn't appear to be very significant for the formation of the DoS, as the formation mechanism is not strongly dependent on the duration of merging. However, if the mass ratio is high, then we might expect the secondary galaxy to contain less dwarfs than the primary galaxy. For mass ratios of 1:2 or higher, the DoS is formed from the secondary galaxy's satellites only. Under these circumstances, if the secondary galaxy contains only a few dwarfs, then the DoS that forms may have few members compared to surrounding dwarfs that are not members. Therefore the closer the mass ratio is to an equal mass merger, the greater the fraction of satellites that can end up in the DoS. Yet a more equal mass ratio will also result in a more major merger which could destroy any pre-existing disk component (i.e. if the primary galaxy were a spiral). This means that, if we could observe DoS in large numbers of galaxies, DoS might be found more preferentially about early type galaxies. The cosmological simulations of \citet{Cautun2015b}, in which large numbers of DoS are found, are dark matter-only, therefore the morphology of their galaxies is not defined. Nevertheless, recent hydrodynamical simulations have shown that the disk component can quickly reform following a merger \citep{Governato2009,Moster2012,Borlaff2014}, so this does not rule out our formation mechanism for DoS seen around disk galaxies. We note that we have simulated a $z$=3 merger, therefore there is sufficient time for a disk to reform. However such a disk might be expected to form with a similar direction of rotation as the orbital plane of the two galaxies, prior to merging. Even if the disk were to reform, it is expected that the merger would enhance the bulge component of the final merger remnant \citep{Stewart2008}. Therefore it would be challenging for our scenario to be the origin of a DoS surrounding a spiral with a low bulge-to-disk mass ratio.

\section{Conclusions}
The aim of this study is to understand if a merger between two galaxies, where one galaxy contains a dwarf galaxy population, could lead to the formation of a Disk of Satellites (DoS). We conduct idealised N-body simulations of an isolated 1:2 mass ratio merger between a primary dark matter halo and a secondary dark matter halo. Within the secondary halo we place a population of 10 dwarf mass dark matter halos. Between simulations, we vary the orbital interaction from plunging to near circular orbits. We also vary the distribution and dynamics of dwarf halos within the secondary halo. For the right type of orbit, and the correct properties of the dwarf halo population, we find we can form DoSs that are large (diameters$\sim$150~kpc) and thin (thickness$\sim$10-40~kpc). These DoSs are formed in the same plane in which the primary and secondary halo interacted (the $x$-$y$ plane). They do not appear through `by-chance' alignments of dwarf halos. In fact, they are generally long lived structures -- they exist from their formation time (after $\sim$0.5~Gyr) for the duration of our simulations (3~Gyr). In fact, we have extended the duration of the standard model simulation and find that its DoS is stable for at least 6~Gyr. The low velocities of the dwarfs out of the plane of the DoS help it to remain thin for longer. The DoS also forms with a clear overall rotation, typically shared by all of the surviving DoS members. The rotation direction matches the direction in which the primary and secondary halo interacted during the merger. In the process of forming the DoS, the dwarf halos population can increase the semi-minor to semi-major axis of their spatial distribution from very thick ($c$/$a$$\sim$0.75) to very thin ($c$/$a$$\sim$0.2). However, this change arises through a large increase in the semi-major axis $a$, and a reduction in the semi-minor axis $c$ (equivalent to the $z$ range).

We test which parameters are important for forming a rotating DoS in our scenario and arrive at the following conclusions.
\begin{enumerate}
\item Prior to the merger, the dwarf population that will end up in the DoS must inhabit a small range of $z$ direction, where $z$ is a vector perpendicular to the orbital interaction plane of the primary and secondary halo. The distribution of the dwarf population in the $x$ and $y$ direction (i.e. radially) is not a strong constraint. 
\item DoS with a large diameter were more readily formed after the merger when the secondary halo enters the primary halo with a roughly circular orbit initially (a tangential velocity from 0.75 to 1.0 times the circular velocity of the two body system). However, even if the orbit is completely circular initially, dynamical friction quickly acts to reduce the initial relative velocity, and induce the merger.
\item When the dwarfs in the secondary halo have initial velocity vectors that are closely aligned with the orbital interaction plane of the primary and secondary halo, in a prograde direction, then the final DoS can remain thin and long-lived ($>$5~Gyr) in our models. In this case, members of the DoS have velocity vectors which end up well aligned with the final DoS plane. Alternatively, if the initial velocity vectors are a mixture of prograde and retrograde, then angular frequency matching between the dwarfs orbiting about the secondary halo, and the secondary halo orbiting about the primary halo, naturally filters the prograde from the retrograde. The prograde dwarfs preferentially end up in the DoS. Meanwhile the retrograde dwarfs end up on highly radial, and tidally destructive, orbits.
\end{enumerate}

We conclude that our scenario for DoS formation could potentially result in a significant fraction of the DoS recently detected in the cosmological simulations of \citet{Cautun2015b}. They estimate that roughly 10$\%$ of galactic halos may contain a DoS, although their properties vary considerably. It is not clear that all of their DoS are long-lived entities. Their condition that satellites share the same direction of rotation does not signify that the satellites will have velocity vectors that are tightly aligned with the plane of the DoS, like in the standard and the thick distribution model. However even to produce relatively short lived rotating DoS in our scenario requires; (i) a small range $z$ range of, at least, some of the dwarfs in a merging halo, and (ii) the galaxies to merge with roughly circular orbits. This double requirement alone may explain why DoS occur in less than one tenth of the galactic halos. To additionally form a long-lived DoS, requires an additional constraint on the dwarf satellite velocity vectors, which likely further reduces their formation probabilility. 

\acknowledgments
RS acknowledges support from Brain Korea 21 Plus Program(21A20131500002) and the Doyak Grant(2014003730). RS
also acknowledges support support from the EC through an ERC grant StG-257720. SKY acknowledges support from the National Research Foundation of Korea (Doyak grant 2014003730). The authors gratefully acknowledge the anonymous referee for constructive comments that improved the paper.

\bibliographystyle{apj}
\bibliography{bibfile}

\clearpage

\end{document}